\documentclass[aps,prl,twocolumn,showpacs,floats,floatfix,superscriptaddress]{revtex4-1}

\usepackage{graphicx}
\usepackage{amsmath}
\usepackage{amssymb}
\usepackage{bm}

\usepackage{color}

\begin{document}

\title{Topologically protected defect states in open photonic systems with non-hermitian charge-conjugation and parity-time symmetry}

\author{Simon Malzard}
\affiliation{Department of Physics, Lancaster University, Lancaster, LA1 4YB, United Kingdom}
\author{Charles Poli}
\affiliation{Department of Physics, Lancaster University, Lancaster, LA1 4YB, United Kingdom}
\author{Henning Schomerus}
\affiliation{Department of Physics, Lancaster University, Lancaster, LA1 4YB, United Kingdom}
\affiliation{Max-Planck-Institut f\"ur Physik komplexer Systeme, 01187 Dresden, Germany}
\date{\today}

\begin{abstract}
We show that topologically protected defect states can exist in open (leaky or lossy) systems even when these systems are topologically trivial in the closed limit.
The states appear from within the continuum, thus in  absence of a band gap, and are generated via exceptional points (a spectral transition that occurs in  open wave and quantum systems with a generalized time-reversal symmetry), or via a degeneracy induced by charge-conjugation-symmetry (which is related to the pole transition of Majorana zero modes).
We demonstrate these findings for a leaking passive coupled-resonator optical waveguide with asymmmetric internal scattering, where the required symmetries (non-hermitian versions of time-reversal symmetry, chirality and charge-conjugation) emerge dynamically.
\end{abstract}
\pacs{03.65.Vf, 42.60.Da, 42.55.Sa, 73.20.At}
\maketitle

Fundamental symmetries appear in a new light when they are discussed within the context of open systems, where particles escape via leakage to the outside world or are absorbed within the material. In these situations one typically encounters decaying normal modes that can be described via the complex eigenfrequencies $\omega_n$ of an effective non-hermitian Hamiltonian, with $H\neq H^\dagger$. This description applies, e.g., on the level of the Helmholtz equation for  dielectric microresonators or photonic crystals, where leaky losses  enter through the boundary conditions while absorption renders the refractive index complex \cite{cao2015}. Such systems still obey reciprocity, $H=H^T$, while the antiunitary time-reversal symmetry $\mathcal{T}H\mathcal{T}= H^*\neq H^T=H$ is in general broken.

There has been much recent interest in open settings where a generalized antiunitary
symmetry $\mathcal{PT}H\mathcal{PT}=H$  still exists \cite{experiment1a,experiment1b,Regensburger,feng2013,eichelkraut2013}.
Here $\mathcal{P}$ stands for parity, in this context understood to be a unitary involution with $\mathcal{P}^2=1$ that is often realized by a geometric  reflection or inversion. True $\mathcal{PT}$ symmetry requires amplifying (active) regions, arranged such that they are mapped via $\mathcal{P}$ onto  corresponding absorptive parts, while leakage needs to be negligible. Eigenfrequencies $\omega_n$ then are either real or occur in a complex-conjugated pair \cite{bender2007}, giving rise to a novel type of mode competition in lasers \cite{schomerus2010,longhi2010,chong2011,feng2014,hodaei2014}. The more general case of a symmetry
\begin{equation}
\mathcal{PT}(H+i\gamma)\mathcal{PT}=H+i\gamma
\label{eq:passivept}
\end{equation}
with a finite offset $\gamma$ also encompasses suitably arranged passive systems where $\mathcal{P}$ now transforms between regions of different losses,  while uniform leakage is also admitted.
The constraints on the spectrum then apply to the shifted frequencies $\Omega_n=\omega_n+i\gamma$, which are either real or appear along with a partner $\Omega_n^*$. Among the many applications, these features can be used, e.g., to engineer band structures in periodic media where the dispersion  is still effectively real or possesses some well-defined additional complex branches \cite{makris,longhi2009,ramezani,tachyons}.

The advent of topological insulators and superconductors \cite{hasan2010,qi2011} has taught us that the classification of universality classes in terms of time-reversal symmetry is incomplete. Two related symmetries, chirality \cite{verbaarschot1993,verbaarschot2000} and charge conjugation \cite{altland1997}, need to be accounted for to identify band structures associated with  finite topological quantum numbers \cite{comment1}, with the most prominent consequence being the formation of spatially localized  defect states at interfaces (points, edges or surfaces) between topologically distinct domains. In particular, a unitary chiral anti-symmetry $\mathcal{X}(H-\Omega^{(0)})\mathcal{X}=-(H-\Omega^{(0)})$ enforces the spectrum to be symmetric around a central frequency $\Omega^{(0)}$, giving rise to frequency pairs $\Omega^{(0)}\pm \Omega_n$. A similar spectral constraint is also enforced by the antiunitary charge-conjugation symmetry $\mathcal{C}=\mathcal{TX}$
that can appear in superconducting systems; this yields pairs $\Omega^{(0)}+ \Omega_n$, $\Omega^{(0)}- \Omega_n^*$ and stabilizes unpaired resonances (broadened Majorana zero modes) at ${\rm Re}\,\Omega_n=0$ \cite{pikulin2012,pikulin2013,sanjose2014}. Within this conventional classification, a necessary requirement for topological nontriviality is the existence of a band gap, into which the defect states then fall. The nascent field of topological photonics \cite{lu2014,raghu2008,wang2008,wang2009,malkova2009,hafezi2011,fang2012,kitagawa2012,khanikev2013,hafezi2013,rechtsman2013a,rechtsman2013b,keil2013,mittal2014,poddubny2014,poli2015} has embarked to realize photonic analogues of these symmetries, while the (beneficial or detrimental?) role of non-hermitian loss and gain has only been considered in settings which are already topological  in the hermitian limit \cite{poli2015,rudner2010,diehl2011,esaki2011,hu2011,HS2013a,HS2013,Zeuner2014}.

Here we identify, for the simple example of a coupled-resonator chain, a mechanism by which topologically protected defect states can appear in open (non-hermitian) systems even when their closed (hermitian) limit is topologically trivial. The defect states form at an interface of two regions with different non-hermiticity, and appear from the continuum of the band structure via two distinct symmetry-protected spectral transitions---where one is associated with $\mathcal{PT}$-symmetry (thus, non-hermitian time-reversal symmetry), while the other is associated with $\mathcal{PC}$-symmetry (thus, the analogously generalized charge-conjugation symmetry that arises from the simultaneous presence of a chiral symmetry). Therefore, robust states as desired for topological photonics can be obtained by combining symmetries with non-hermitian effects that go beyond the universal classification of electronic systems.

\begin{figure}
\includegraphics[width=\columnwidth]{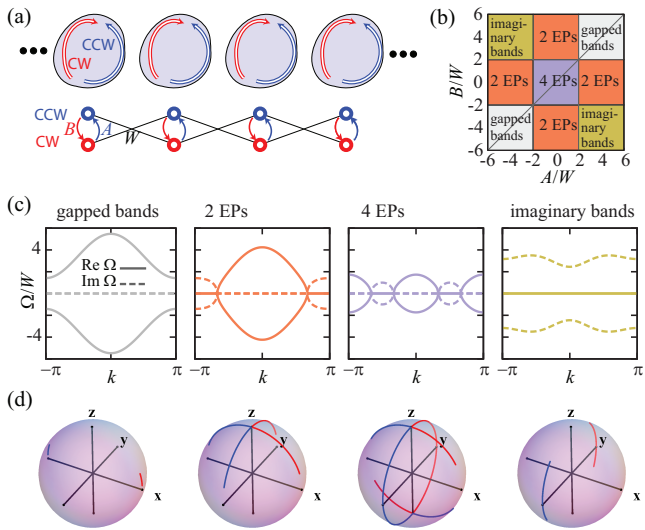}
\caption{\label{fig1} (color online) (a) Coupled-resonator waveguide with internal asymmetric scattering (couplings $A$ and $B$) between a counterclockwise (CCW) and a clockwise (CW) wave component, and coupling $W$ between CCW and CW components in neighboring resonators. For real couplings this system realizes non-hermitian versions of time-reversal symmetry, chirality and charge conjugation. (b) Phase diagram for the bulk dispersion \eqref{eq:dispall}. The dispersion can be real and gapped, exhibit 2 or 4 exceptional points (EPs) at which real and imaginary branches meet, or be fully imaginary. In the examples in (c),
$A/W=4$ and $B/W=3$ (real gapped dispersion), $A/W=4$ and $B/W=1$ (2EPs), $A/W=1$ and $B/W=-1$ (4EPs), as well as $A/W=4$ and $B/W=-3$ (imaginary dispersion). (d) Corresponding representation of the complex bands on the Bloch sphere (upper band $\Omega_+$, red; lower band $\Omega_-$, blue).}
\end{figure}

{\em Dynamical realization of non-hermitian parity-time and charge-conjugation symmetry.---}
We first describe how the required symmetries can be implemented in a simple passive resonator chain, with losses solely provided due to leakage but without any need of absorption or amplification (which then also translates to analogous open quantum systems).  This can be achieved in a coupled-resonator optical waveguide (CROW)  \cite{little1997,stefanou1998,yariv1999} which consists of identical asymmetric cavities \cite{wiersig2008}, as sketched in Fig.~\ref{fig1}(a).
Each individual resonator features two modes---a counterclockwise  (CCW) propagating mode with amplitude $a_n$ and a clockwise (CW) propagating mode with amplitude $b_n$, which we group into a vector $\psi_n=(a_n,b_n)^T$.
Resonator arrays in which these two modes are well decoupled feature in setups that realize photonic topological edge states in analogy to the quantum-Hall effect
\cite{hafezi2011,hafezi2013}. In our setting, however, the internal coupling between these modes is desired, and the key feature is that this coupling can be made asymmetric by opening the system, even if no magnetic field is applied---as has been established in recent works on individual resonators \cite{wiersig2008,wiersig2011b,wiersig2011,yi2011,scott2012,wiersig2014}.
The coupling of the modes is then described by a nonhermitian internal Hamiltonian \cite{wiersig2008,wiersig2014}
\begin{equation}
h=\left(\begin{array}{cc}\Omega^{(0)} & A \\B & \Omega^{(0)}\end{array}\right),
\end{equation}
where the
constants $A\neq B^*$ and $\Omega^{(0)}$ account for the asymmetric internal scattering and the losses within the cavity.
Throughout the chain, the coupling between adjacent resonators is dominantly between CCW and CW waves, so that the coupling matrix is \cite{wiersig2014,HS2014}
\begin{equation}
t=\left(\begin{array}{cc}0 & W\\W & 0\end{array}\right).
\end{equation}
In coupled-mode approximation, the stationary wave equation then takes the form
\begin{equation}
\omega \psi_n=h\psi_n+t(\psi_{n+1}+\psi_{n-1}),
\end{equation}
which admits Bloch solutions $\psi_n=\exp(ikn)\Psi$.
The associated Bloch Hamiltonian is
\begin{equation}
h(k)=\left(\begin{array}{cc}\Omega^{(0)} & A+2W\cos(k) \\B+2W\cos(k) & \Omega^{(0)}\end{array}\right),
\end{equation}
and leads to the dispersion relation $\omega_\pm(k)=\Omega^{(0)}+\Omega_\pm(k)$,
\begin{equation}
\Omega_\pm(k)=\pm\sqrt{(A+2W\cos k)(B+2W\cos k)},
\label{eq:dispall}
\end{equation}
where the subscript $\pm$ labels two bands. The symmetry about $\Omega^{(0)}$ is a consequence of a chiral symmetry with $\mathcal{X}\psi_n =\sigma_z\psi_n$, which maps the bands onto each other. The chiral symmetry is thus realized by the freedom of the relative sign of the CCW and CW amplitudes (a gauge freedom compatible with time-reversal symmetry which generically appears in systems with two mutually coupled sublattices).

As shown by exact numerical calculations \cite{wiersig2014,HS2014}, for representative resonator geometries $A$, $B$ and $W$ are almost real, and can be further tuned towards real values by adjustments of a few shape parameters. We thus neglect the imaginary parts of these parameters.
Apart from an offset $\gamma=i\,{\rm Im}\,\Omega^{(0)}$, the dispersion is then either real (in some range of $k$) or purely imaginary (in the complementary range of $k$).
These $k$ ranges are joined by degeneracies, known as exceptional points \cite{kato,heiss2000,heiss2012,berry2004,wiersig2014b},
where $\omega_\pm(k)=\Omega^{(0)}$, thus $\cos k=-A/2W$ or $\cos k=-B/2W$.
A completely real dispersion with a gap is achieved if
$|A/2W|>1$ and $|B/2W|>1$, provided $AB>0$ [see the phase diagram in Fig.~\ref{fig1}(b) and representative dispersions in Fig.~\ref{fig1}(c)].

The underlying symmetry can be made explicit by a basis change, $\phi_n=2^{-1/2}(i\sigma_x+\openone)\psi_n$ where $\sigma_x$ is a Pauli matrix,
after which the Bloch Hamiltonian takes the form
\begin{equation}
\tilde h(k)=\frac{1}{2}\left(\begin{array}{cc}2\Omega^{(0)}+i(A-B) & A+B +4W\cos k \\A+B  +4W\cos k & 2\Omega^{(0)}-i(A-B)\end{array}\right)
.
\end{equation}
The Hamiltonian is now symmetric, as required by reciprocity (which is hidden in basis of CCW
and CW modes since propagating waves are complex), and furthermore exhibits a passive $\mathcal{PT}$ symmetry \eqref{eq:passivept} with $\mathcal{P}=\sigma_x$, and $\gamma=-{\rm Im}\,\Omega^{(0)}$. The chiral symmetry is transformed to $\mathcal{X}=\sigma_y$, and commutes with $\mathcal{PT}$. This realizes all the symmetries mentioned in the introduction, including $\mathcal{PC}=\mathcal{PTX}$, with respect to the central frequency $\Omega^{(0)}$. From here on, we work in terms of the shifted frequencies $\Omega=\omega-\Omega^{(0)}$, for which the dispersion is directly given by  Eq.~\eqref{eq:dispall}.

Having established these symmetries we now return to the basis of CCW
and CW modes and discuss topological aspects of the band structure. For this we consider the $k$ dependence of the Bloch vectors $\Psi(k)$, which we interpret as pseudospins with polarization vector $\vec P=\langle(\sigma_x,\sigma_y,\sigma_z)\rangle$.
In the hermitian limit $B=A$ (both real), the two bands $\Omega_\pm(k)=\pm (A+2W\cos k)$ arise from $k$-independent pseudospins $\Psi_\pm=2^{-1/2}(1,\pm 1)^T$, with $\vec P=(\pm 1,0,0)$ pointing along the $x$ axis. The absence of any winding of the pseudospin renders the system topologically trivial, so that we do not expect any defect states in the presence of interfaces, even if there is a gap.
In the non-hermitian case, we can write $\Psi_\pm(k)\propto(A+2W\cos k,\Omega_\pm(k))^T$. As shown in Fig.~\ref{fig1}(d), the polarization vector now acquires $k$ dependence; it is confined to the $xz$ plane when the dispersion is real and to the $yz$ plane when the dispersion is imaginary. These branches are again joined at the exceptional points, where $\vec P$ points up or down along the $z$ axis, with  $\vec P_-(k)=R_z(\pi)\vec P_+(k)$ related by a $\pi$ rotation about the $z$ axis. In particular, the way these points are connected depends on whether $A>B$ or $A<B$ (with the two cases related by a rotation $R_x(\pi)$ by $\pi$ about the $x$ axis). Does the system now admit defect states?

\begin{figure}
\includegraphics[width=\columnwidth]{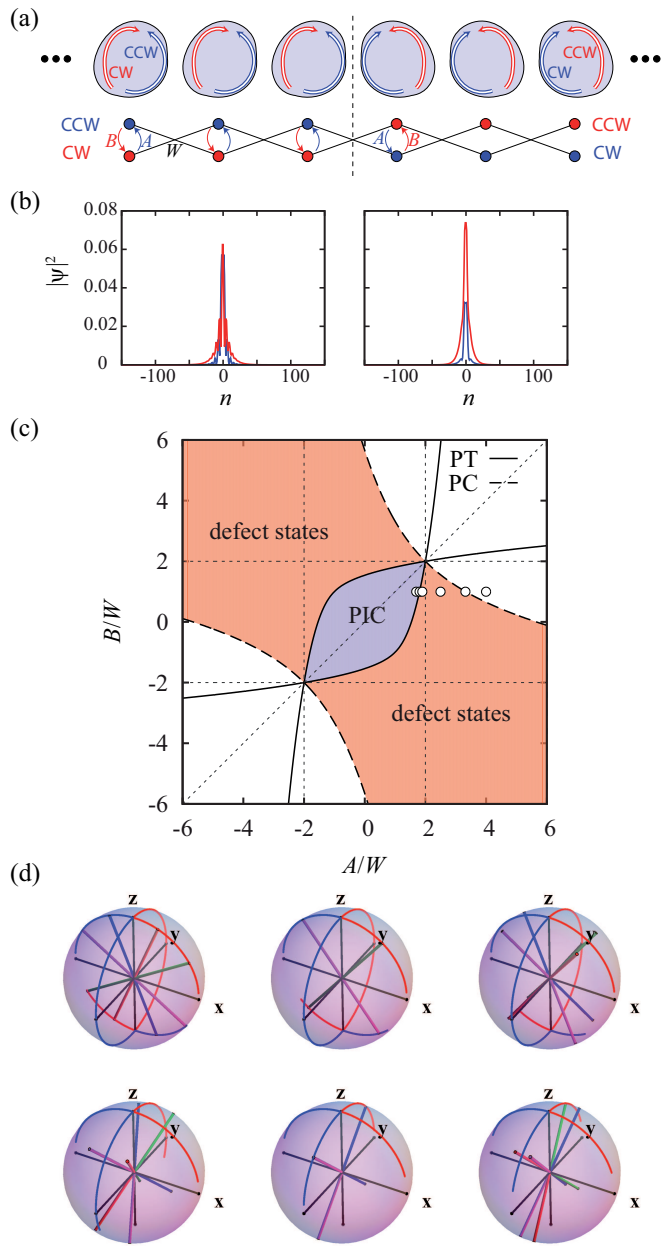}
\caption{\label{fig2} (color online)
(a) Coupled-resonator waveguide with a defect, created by inverting the orientation
of the resonators in half of the system. In the closed limit, the system is trivial, and the defect does not create any bound states. (b) Defect states in a system of 300 resonators, with  $A/W=1.9$ (left panel) and $A/W=2.5$ (right panel), while $B/W=1$.
(c) Phase diagram indicating the existence of defect states, as well as their extended-state precursors (realizing perfect interband transitions, PIC).  The boundaries of the defect phase are given by degeneracy conditions. At the PT boundary, the extended PIC states bifurcate into pairs $\Omega_n$, $\Omega_n^*$ of defect states that are related by the $\mathcal{PT}$ symmetry. At the PC boundary one encounters a degeneracy of charge-conjugated partner states $\Omega_n$, $-\Omega_n^*$, beyond which the defect states are non-normalizable. (d) Bloch-sphere position of the defect states (rods) relative to the bulk dispersion (lines) [parameters $A/W=1.7$, 1.81 (PT), 1.9, 2.5, 3.32 (PC), 4 with $B/W=1$, as indicated by the white circles in (c)].}
\end{figure}

{\em Defect states.---}
In order to answer this question, we create a defect in the chain by inverting the orientation of the resonators in half of the system [see Fig.\ \ref{fig2}(a)]. From the traditional perspective of  hermitian systems, the defect cannot be classified as topological, and does not give rise to any defect states. In the non-hermitian setting, we will see that the defect acquires topological features in a spectral phase transition at which localized defect states emerge [as illustrated in Fig.\ \ref{fig2}(b)]. The phase transition
takes the form of a $\mathcal{PT}$-induced exceptional point along one part of the phase boundary, while it is associated with a $\mathcal{PC}$-induced degeneracy along the other parts of the phase boundary [this is summarized in Figs.~\ref{fig2}(c,d), to which we refer throughout the remaining discussion].

It is easy enough to identify the conditions for the formation of defect states.
In the presence of the defect, the wave equation takes the form
\begin{equation}
\Omega\psi_n=h_{n}\psi_n+t(\psi_{n+1}+\psi_{n-1}),
\label{eq:waveequation}
\end{equation}
where now $h_n=(h-\Omega^{(0)})$ for $n<0$ (left half of the chain) and $h_n=(h-\Omega^{(0)})^T$ for $n\geq0$ (right half of the chain). At any fixed $\Omega$, in each half of the system the solutions are still obtained from a superposition of Bloch waves with
\begin{align}
2W\cos k_\pm&=c_\pm=-\frac{A+B}{2}\pm\sqrt{\frac{(A-B)^2}{4}+\Omega^2},
\end{align}
associated with pseudospins
\begin{equation}
\Psi^{(L)}_\pm(\Omega)\propto\left(\begin{array}{c}A+c_\pm\\ \Omega\end{array}\right), \quad \Psi^{(R)}_\pm(\Omega)\propto\left(\begin{array}{c}\Omega \\ A+c_\pm\end{array}\right).
\label{eq:defectvecs}
\end{equation}
Each of the values $c_+$, $c_-$ are associated with a pair of Bloch waves with propagation factors $\exp(ik_\pm)=\lambda_\pm$, $\exp(-ik_\pm)=(\lambda_\pm)^{-1}$,
where we choose $k_\pm$ such that $|\lambda_\pm|\geq 1$ if $k_\pm$ is complex.
We now match the solutions with propagation factor $\exp(ik_\pm)$ in the left part to the solutions with propagation factor $\exp(-ik_\pm)$ in the right part.
One then finds the condition
\begin{equation}
(\Omega-2W)^2\Omega -A B\Omega=(A-B)^2 W/2
\label{eq:sym}
\end{equation}
for defect states with a symmetric wavefunction, and
\begin{equation}
-(\Omega+2W)^2\Omega +A B\Omega=(A-B)^2 W/2
\label{eq:antisym}
\end{equation}
for defect states with an antisymmetric wavefunction.
Because of the $\mathcal{PT}$ symmetry, solutions are again either real or appear in complex conjugated pairs. Furthermore, chirality maps $\Omega\to -\Omega$, corresponding to a transformation between symmetric and antisymmetric wavefunctions.
Thus, the defect states come in a quadruple of frequencies $\Omega_n$, $\Omega_n^*$, $-\Omega_n$, $-\Omega_n^*$. The condition that the corresponding wave function indeed decays leads to the phase diagram in Fig.\ \ref{fig2}(c).

In this phase diagram, the hermitian case $A=B$ defines a diagonal. The defect states are confined to a region away from the diagonal, which is bounded by two different transitions. Along the curves labeled PT, where
$27(A + B)^4=16 A^2 B^2 (1+ AB/W^2) + 8(8W^2+9AB) (A + B)^2$,
a pair of real solutions $\Omega_n$ of
Eq.~\eqref{eq:sym} bifurcates into a pair of complex-conjugated solutions.
Before this exceptional point, the solutions are real, with $|\lambda_\pm|=1$, and describe the scattering of an incoming extended state in one band into an outgoing extended state in the other band. This region of perfect interband conversion is labeled PIC. At the exceptional point, the propagation factors $\lambda_+$ of the two solutions coalesce, and so do the factors $\lambda_-$; beyond the exceptional point we then have $|\lambda_\pm|>1$, giving rise to properly normalizable defect states. The same scenario occurs simultaneously for the chirality-related solutions of Eq.~\eqref{eq:antisym}.
The second kind of transition appears along the curves labeled PC, where $A^2+6AB+B^2=32W^2$. There, a complex solution $\Omega_n$ of Eq.~\eqref{eq:sym} coalesces with a charge-conjugated solution $-\Omega_n^*$
of Eq.~\eqref{eq:antisym}, meaning that they are purely imaginary. This is similar to the pole transition of broadened Majorana zero-modes, which are then pinned to the imaginary axis and become their own charge-conjugated partner.
These transitions also occur in skew-Hamiltonian ensembles governing the topological transitions in Josephson junctions \cite{beenakker2013}. In the present problem, the PC transition signals the point where the matching conditions can only be fulfilled by combining decaying with increasing wavefunctions, which occurs when one of the wave numbers $k_+$, $k_-$ crosses the real axis.

At both types of transition, the defect states therefore interact with the real branch of the dispersion relation $\Omega_\pm(k)$ (for the exceptional points along the PT boundary), or with the purely imaginary branch of this dispersion relation (for the charge-conjugation-induced degeneracy along the PC boundary). On the level of the wavefunctions, this interaction is again revealed via the corresponding polarization vectors. Focussing on the wave function in the left part of the system, we have
\begin{align}
\vec P^{(L)}_-(\Omega)=-\vec P^{(L)}_+(\Omega^*)
\end{align}
for the two partial waves $\Psi_\pm^{(L)}$ given in Eq.~\eqref{eq:defectvecs}, while the $\mathcal{PT}$ and chiral symmetries relate
\begin{align}
\vec P^{(L)}_\pm(\Omega^*)&=-R_y(\pi)\vec P^{(L)}_\pm(\Omega),\\
\vec P^{(L)}_\pm(-\Omega)&=R_z(\pi)\vec P^{(L)}_\pm(\Omega),
\end{align}
where $R_a$ denotes a $\pi$ rotation about axis $a$.
[We also have $\vec P^{(R)}_\pm(\Omega)=R_x(\pi)\vec P^{(L)}_\pm(\Omega)$.]
In Fig.\ \ref{fig2}(d), we show how these polarization
vectors interact.
In the PIC phase, each state corresponds to a pair of opposite vectors $\vec P^{(L)}_\pm$ confined to the $xz$ plane (the locus of the real dispersion branch), with chirality-related partner states connected by a $\pi$ rotation about the $z$ axis.
At the PT transition, the vectors bifurcate and move out of the $xz$ plane. In the defect phase, the two vectors $\vec P^{(L)}_\pm$ for a given defect state make an angle, but remain related by a $\pi$ rotation about the $y$ axis. The partner state with frequency $\Omega^*$ points into the opposite direction, while the chirality-related states are still obtained by a $\pi$ rotation about the $z$ axis. At the PC transition, each state collides with a charge-conjugated partner at a point in the $yz$ plane  (the locus of the imaginary dispersion branch). These interactions all occur at symmetry-protected positions, which renders the defect phase topologically stable. For numerical verification of this robustness in finite and disordered systems see \cite{supmat}.

In summary, robust defect states can exist in open systems that are topologically trivial in the closed limit. We illustrated this for a leaky optical resonator chain where  defect states appear at an interface between regions in which hermiticity is broken in different ways (in contrast, non-hermiticity is not sufficient to create edge states at the end of a finite sample).
The states are topologically protected as they arise in spectral phase transitions that are linked to the spontaneous breaking of fundamental symmetries (parity-time symmetry and charge-conjugation symmetry) for a sufficient degree of non-hermiticity in the system.
The required symmetries are realized when the couplings in the propagating wave basis are real. As this does not require any absorption, the formation mechanisms described here  also translate to analogous geometrically open quantum systems, including electronic systems which are suitably coupled to external reservoirs. Our observations raise new questions, such as whether it is possible to characterize these systems in terms of topological quantum numbers, and more generally whether they can be understood by a suitable extension of the conventional topological classification of closed systems.

\begin{acknowledgments}
We thank Jan Wiersig for useful discussions and comments.
This research was supported by EPSRC via grant EP/J019585/1.
The data created during this research is openly available \cite{rdm}.

\end{acknowledgments}


\makeatletter
\renewcommand{\theequation}{A.\arabic{equation}}
\renewcommand{\thefigure}{A\@arabic\c@figure}
\makeatother
\setcounter{equation}{0}
\setcounter{figure}{0}

\appendix

\section{Appendix: Finite systems and disorder}

To illustrate the robustness of the defect states and the PT and PC phase transitions, we extend our considerations to finite systems as well as situations in which $A$ and $B$ are subject to further asymmetries and disorder, with different fixed or  average values in the left and right halves of the system.  The left-right symmetry (which we did not exploit in our general considerations) is then broken, while the $\mathcal{PT}$ and $\mathcal{PC}$ symmetries continue to hold in any given system.
To facilitate this discussion, we first introduce a number of useful quantities to characterize and identify the defect states in such general settings.

\begin{figure}[t]
\includegraphics[width=\columnwidth]{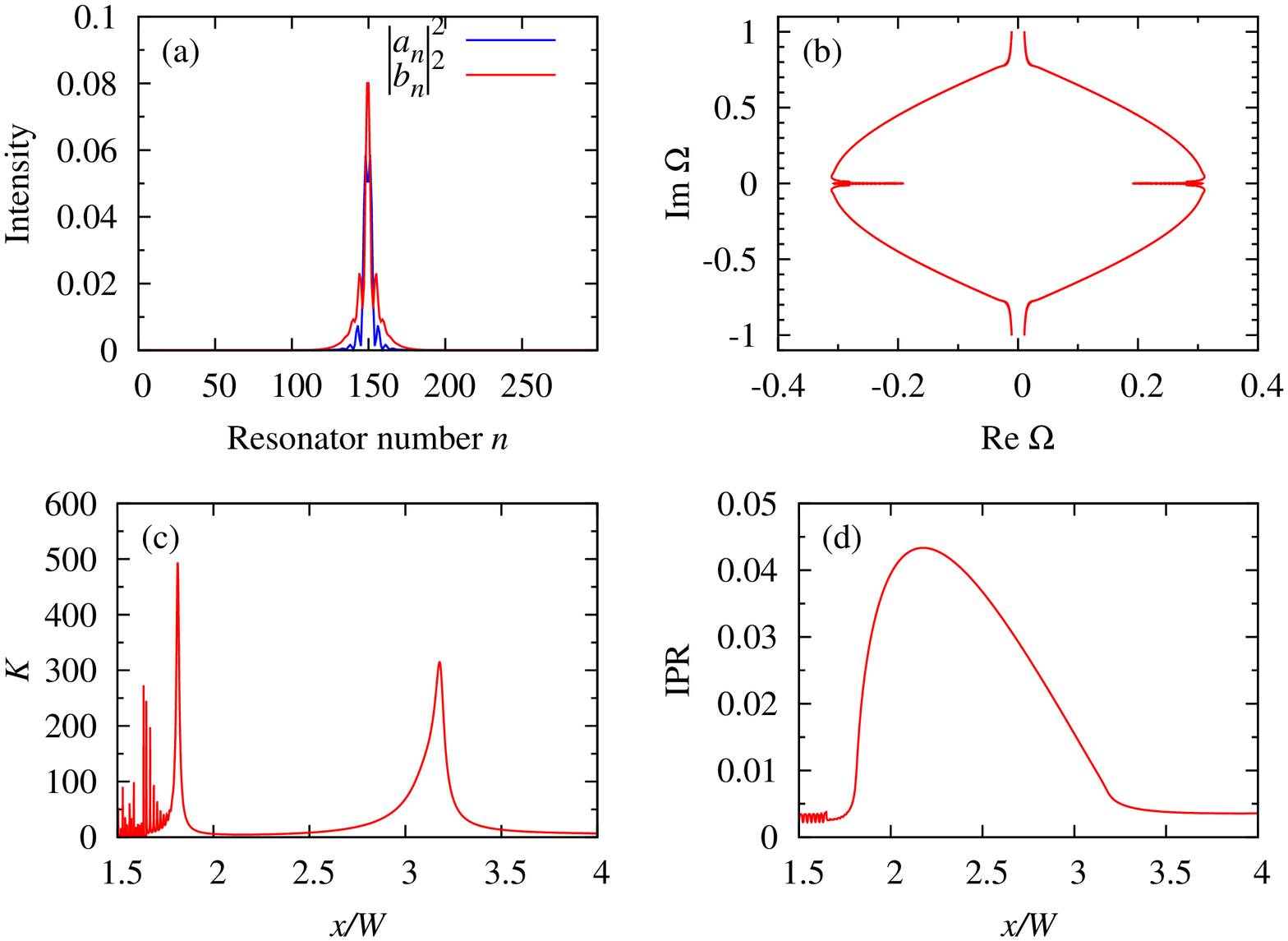}
\caption{\label{fig3} (color online) (a) Mode profile of a defect state in a finite system consisting of 300 asymmetric resonators with $A=W$ and $B=2\,W$  [blue: intensity $|a_n|^2$ of the CCW component in the left half and CW component in the right half; red:  intensity  $|b_n|^2$ of the CW component in the left half and CCW component in the right half].
Panel (b) shows the trajectories of the eigenvalues associated with the defect states and their predecessors as
the value of $B=x$ is changed over the range $1<x<4.5$ at fixed $A=W$.
(c) Petermann factor $K$, Eq.~\eqref{eq:pf}, as a function of $x$.
The Petermann factor becomes large at the  PT and PC phase boundaries, as well as at the exceptional points in the PIC phase, in accordance with the spectral degeneracies encountered there. (d) Inverse participation ratio (IPR),
Eq.~\eqref{eq:ipr}, showing strong localization of the defect state in the expected range.
 }
\end{figure}

Figure ~\ref{fig3}(a) shows a defect state in a finite system of 300 resonators, with $A=W$ and $B=2\,W$. The defect is placed into the center, so that there are two half chains of 150 resonators.
The amplitude $a_n$ refers to the CCW component in the left half of the chain and the CW component in the right half of the chain, while  $b_n$ refers to the CW component in the left half of the chain and the CCW component in the right half of the chain; see coloring of sites in Fig.~2
(a) in the main text.

By changing the couplings smoothly according to $B=x$, we can steer the system from the defect phase over the PT and PC phase boundaries. The trajectories of the quadruple of defect eigenvalues and their precursors are shown in  Fig.~\ref{fig3}(b).
At the PT transition, we find that the  degeneracy on the real axis is exactly realized even in the finite system. We relate this to the fact that in the wave matching problem, the defect state then morphs into the PIC precursor state, which simply is a specific example of an extended state.
At the PC transition,  in the wave-matching picture the defect state morphs into a non-normalizable state. As the spectrum of the finite system cannot lose an eigenvalue, the exact degeneracy on the imaginary axis is then lifted; we find that the state hybridizes with extended states as soon as its decay length becomes comparable with the system size.
The exact degeneracy is then approached more and more closely when the system size is increased.

In Fig.~\ref{fig3}(c), we quantify the closeness to spectral degeneracy by the Petermann factor
\begin{equation}
K=\frac{\langle L|L\rangle \langle R|R\rangle}{|\langle L|R\rangle|^2}.
\label{eq:pf}
\end{equation}
This is a measure of  non-orthogonality of the left eigenstates $\langle L|$ and right eigenstates $| R\rangle$ which diverges at non-hermitian degeneracies \cite{berry2004,wiersig2014b} (note that the notion of left and right eigenstates does not refer to the left and right half of the chain, but to the left and right multiplication of the states with the non-hermitian effective Hamiltonian).  The Petermann factor also quantifies the sensitivity of the system to general perturbations \cite{wiersig2014b}, including quantum noise \cite{HS2000,HS2009}, which thus can be used to map out the boundaries of the defect phase in such general settings.
We find that the Petermann factor becomes large at the  PT and PC phase boundaries, in accordance with the spectral degeneracies that one approaches there.

In order to assess the localization of the defect state  we use the inverse participation ratio
\begin{equation}
{\rm IPR}=\sum_n(|a_n|^4+|b_n|^4),
\label{eq:ipr}
\end{equation}
which is shown in Fig.~\ref{fig3}(d). As expected, the IPR is large within the defect phase.

\begin{figure}[t]
\includegraphics[width=\columnwidth]{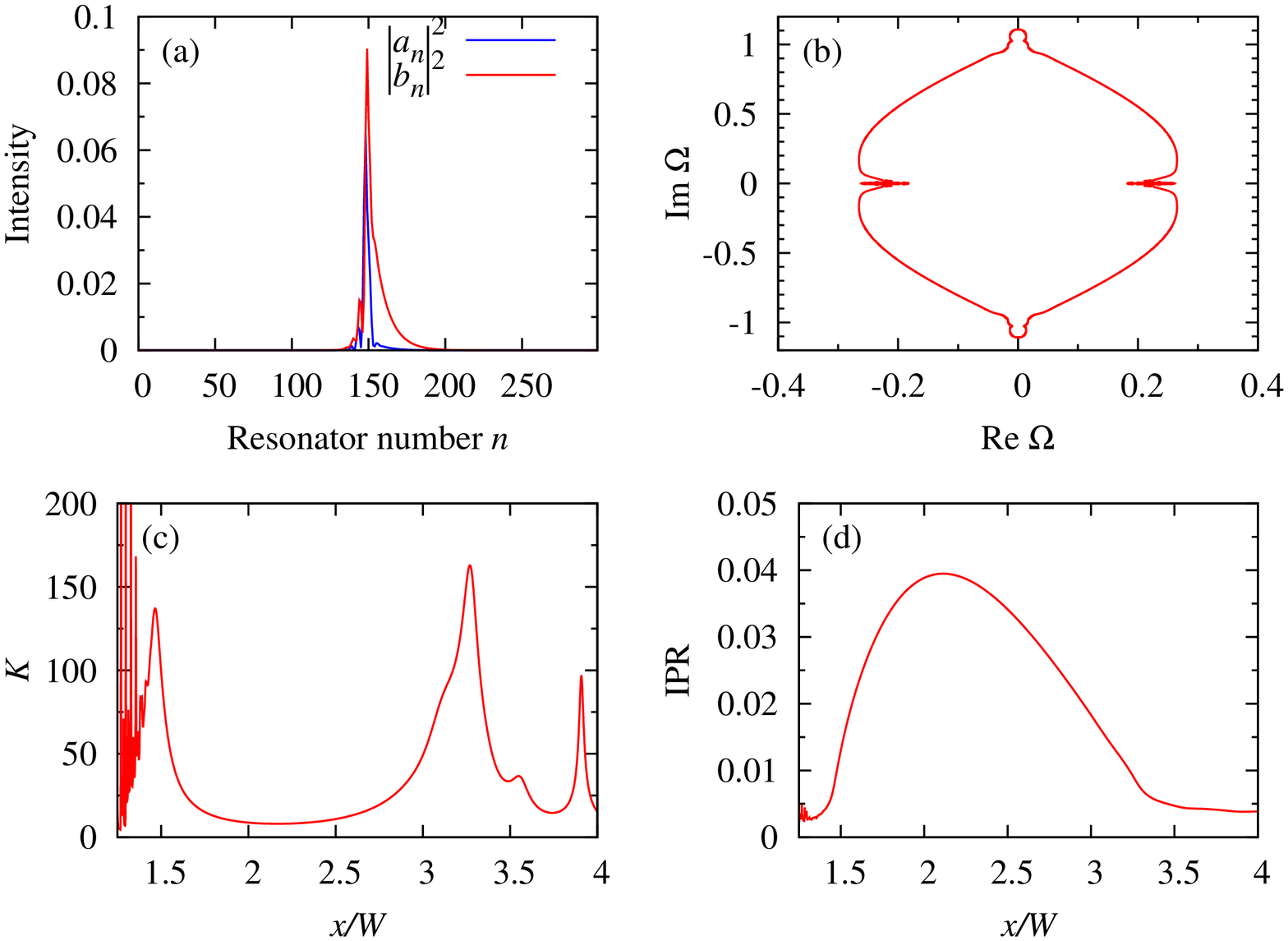}
\caption{\label{fig4} (color online)
Same as Fig.~\ref{fig3}, but for couplings $A=W$ and $B=2\,W$ in the left half and $A=0.5\,W$, $B=2.5\,W$ in the right half of the chain (panel a), as well as $A=W$, $B=x$ (left half) and $A=0.5\,W$, $B=x+0.5\,W$ (right half) in the remaining panels.
}
\end{figure}

\begin{figure}[t]
\includegraphics[width=\columnwidth]{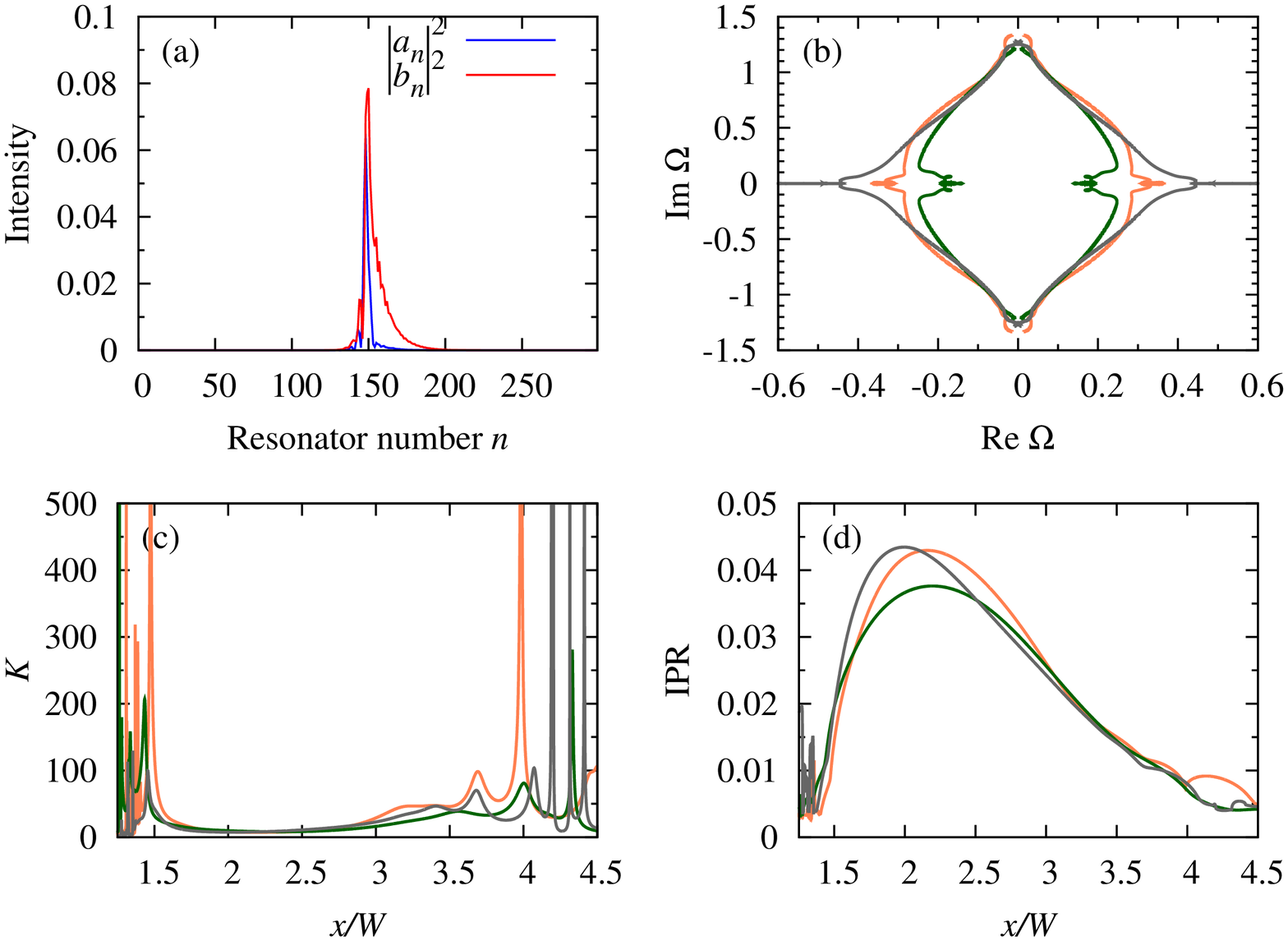}
\caption{\label{fig5} (color online)
Same as Fig.~\ref{fig4}, but with the couplings $A$ and $B$ in each individual resonator perturbed by
independent amounts $y$, drawn from a uniform box distribution $y\in[-0.1\,W,0.1\,W]$. The results represent one disorder realization in panel (a), while the other panels compare the results for three disorder realizations.
}
\end{figure}


As shown in Fig.~\ref{fig4}(a) for an asymmetric system  ($A=W$ and $B=2\,W$ in the left half and $A=0.5\,W$, $B=2.5\,W$ in the right half of the chain), well-defined defect states still exist in this more general situation, where we now observe a different decay of the intensity into the two half chains. By changing the couplings smoothly, now according to $B=x$ (left half) and $B=x+0.5\,W$ (right half), we again find that the defect phase terminates at PT and PC transitions, where the Petermann factor becomes large. The IPR is again large within the defect phase.

Figure~\ref{fig5} displays the same information in the presence of additional disorder, where the couplings $A$ and $B$ in each resonator are modified by independent perturbations $y$ from a uniform distribution $y\in[-0.1\,W,0.1\,W]$. This breaks the translational invariance in each half chain an induces Anderson localization, with a localization length that competes with the defect state, especially when the boundaries of the defect phase are approached.
While the disorder shifts the PT and PC transitions in parameter space, the defect phase remains well defined, and the features of the defect states are remarkably robust.

\end{document}